\documentstyle[preprint,aps,psfig]{revtex}
\newcommand{\beq}{\begin{equation}}
\newcommand{\eeq}{\end{equation}} 
\newcommand{\beqa}{\begin{eqnarray}}
\newcommand{\eeqa}{\end{eqnarray}}
\newcommand{\ba}{\begin{array}}
\newcommand{\ea}{\end{array}}

\begin{document}
\draft

\widetext 
\title{Mean-Field vs Monte-Carlo equation of state 
for the expansion of a Fermi superfluid in the BCS-BEC crossover} 
\author{L. Salasnich$^{1,2}$ and N. Manini$^{2}$} 
\address{$^{1}$CNISM and CNR-INFM, Unit\`a di Padova, \\ 
Dipartimento di Fisica ``Galileo Galilei'', 
Universit\`a di Padova, \\
Via Marzolo 8, 35122 Padova, Italy \\
$^{2}$Dipartimento di Fisica and CNISM,
Universit\`a di Milano, \\
Via Celoria 16, 20133 Milano, Italy} 

\maketitle

\begin{abstract} 
The equation of state (EOS) of a Fermi superfluid is investigated 
in the BCS-BEC crossover at zero temperature. 
We discuss the EOS based on Monte-Carlo (MC) data and 
asymptotic expansions and the EOS derived from the 
extended BCS (EBCS) mean-field theory. Then we introduce 
a time-dependent density functional, based on the bulk EOS 
and Landau's superfluid hydrodynamics with a 
von Weizs\"acker-type correction, 
to study the free expansion of the Fermi superfluid. 
We calculate the aspect ratio and the released energy 
of the expanding Fermi cloud showing that MC EOS and EBCS EOS 
are both compatible with the available 
experimental data of $^6$Li atoms. 
We find that the released energy satisfies an approximate 
analytical formula that is quite accurate in the BEC regime. 
For an anisotropic droplet, our numerical simulations 
show an initially faster reversal 
of anisotropy in the BCS regime, later suppressed 
by the BEC fluid. 
\end{abstract}

\narrowtext 

\newpage 

\section{Introduction} 

Current experiments with a Fermi gas 
of $^{6}$Li or $^{40}$K atoms in two hyperfine spin states  
operate in the regime of deep Fermi degeneracy. 
The experiments are concentrated across a Feshbach resonance, 
where the s-wave scattering length $a_F$ of the interatomic 
Fermi-Fermi potential 
varies from large negative to large positive values. 
In this way it has been observed a crossover from a 
Bardeen-Cooper-Schrieffer (BCS) 
superfluid to a Bose-Einstein condensate (BEC) 
of molecular pairs \cite{duke,ens,mit}. 
\par
The bulk energy per particle 
of a two-spin attractive Fermi gas can be expressed 
\cite{duke,ens,mit,perali,mc1}
in the BCS-BEC crossover by the following equation 
\beq 
{\cal E}(n) = {3\over 5} \; 
{\hbar^2 k_F^2\over 2m} \; f(y) \; , 
\eeq
where $k_F=(3\pi^2 n)^{1/3}$ is the Fermi wave vector, 
$n$ is the number density, 
and $f(y)$ is a universal function of the inverse 
interaction parameter $y=(k_F a_F)^{-1}$, with $a_F$ 
the Fermi-Fermi scattering length. The full behavior of the 
universal function $f(y)$ is unknown but 
one expects that in the BCS regime ($y\ll -1$) it has 
the following asymptotic behavior 
\beq 
f(y)=1+{10\over 9\pi\, y} + O({1\over y^2}) \; ,   
\eeq 
as found by Yang {\it et al.} \cite{yang1,yang2} in 1957. 
In this regime 
the system is a Fermi gas of weakly bound Cooper pairs where the 
superfluid gap energy $\Delta$ is exponentially small. 
Instead, in the unitarity limit ($y=0$) 
the energy per particle is proportional to that 
of a non-interacting Fermi gas and, 
from Monte-Carlo (MC) results \cite{mc1}, one finds 
\beq 
f(0)=0.42 \pm 0.02\; . 
\eeq
Finally, in the BEC regime ($y\gg 1$) 
the system is a weakly repulsive Bose gas of molecules 
of mass $m_M=2 m$, density $n_M=n/2$ and interacting  
with $a_M=0.6 a_F$ (from MC results \cite{mc1} and 4-body theory 
\cite{russi}). 
In this BEC regime one expects the asymptotic expression 
\beq 
f(y)= {5 a_M\over 18 \pi a_F\, y} + O({1\over y^{5/2}}) \; ,    
\eeq 
as found by Lee, Yang and Huang \cite{lee}, again in year 1957. 

\section{Monte-Carlo vs Mean-field}

We have recently shown \cite{nick1} that the unknown universal 
function $f(y)$ can be modelled by the analytical formula 
\beq 
f(y) = \alpha_1 - \alpha_2 
\arctan{\left( \alpha_3 \; y \; 
{\beta_1 + |y| \over \beta_2 + |y|} \right)}  
\label{f-ms} \; . 
\eeq
This formula has been obtained 
from Monte-Carlo (MC) simulations \cite{mc1} 
and the asymptotic expressions. Table 1 of Ref. \cite{nick1}  
reports the values of the interpolating value of 
$\alpha_1$, $\alpha_2$, $\alpha_3$, $\beta_1$ 
and $\beta_2$.  The thermodynamical formula 
\beq 
\mu (n) = {\partial \left( n {\cal E}( n ) \right) 
\over \partial n } = {\hbar^2 k_F^2\over 2m} 
\left( f(y) - {y \over 5} f'(y) \right) \; .  
\label{mu-ms} 
\eeq
relates the bulk chemical potential $\mu$ to the 
energy per particle ${\cal E}$.
We call Monte-Carlo equation of state (MC EOS)  
the equation of state $\mu=\mu(n,a_F)$ 
obtained from Eqs. (\ref{f-ms}) and (\ref{mu-ms}). 
\par 
Within the mean-field theory, the chemical 
potential $\mu$ and the gap energy $\Delta$ of the uniform Fermi 
gas are instead found by solving the following 
extended BCS (EBCS) equations \cite{nick2,marini} 
\beq 
-{1\over a_F} = {2 (2m)^{1/2} \over \pi \hbar^3} \,
\Delta^{1/2} \, 
\int_0^{\infty} dy \, y^2 \,
\left(
{1\over y^2} - {1\over \sqrt{(y^2-{\mu\over \Delta})^2+1} }
\right) \, , 
\label{ebcs1} 
\eeq
\beq 
n = {N\over V} = {(2m)^{3/2} \over 2 \pi^2 \hbar^3} \,
\Delta^{3/2} \, 
\int_0^{\infty} dy \, y^2 \,
\left(
1 - {(y^2-{\mu\over \Delta})
\over \sqrt{(y^2-{\mu\over\Delta})^2+1} }
\right) \; . 
\label{ebcs2} 
\eeq 
By solving these two EBCS equations 
one obtains the chemical potential $\mu$ as a 
function of $n$ and $a_F$. 
Note that EBCS theory does not predict the correct BEC limit: 
the molecules have scattering length $a_M=2a_F$ instead of 
$a_M=0.6a_F$. We call EBCS equation of state 
(EBCS EOS) the mean-field equation of state $\mu=\mu(n,a_F)$ 
obtained from Eqs. (\ref{ebcs1}) and (\ref{ebcs2}). 
Obviously, our MC EOS is much closer than the EBCS EOS 
to the MC results obtained in Ref. \cite{mc1} 
with a fixed node technique. 
\par 
For completeness, 
we observe that within the EBCS mean-field theory  
the condensate density $n_0$ of the Fermi superfluid 
can be written in terms of a simple formula 
\cite{nick2,ortiz}, given by 
\beq 
n_0 = {m^{3/2} \over 8 \pi \hbar^3} \,
\Delta^{3/2} \sqrt{{\mu\over \Delta}
+\sqrt{1+{\mu^2 \over \Delta^2} }} 
\label{con0} 
\; .   
\eeq 
In Ref. \cite{nick2} we have found that the condensate 
fraction is exponentially small 
in the BCS regime ($y\ll -1$) and goes to unity
in the BEC regime ($y\gg 1$). A very recent MC calculation 
\cite{mc2} has confirmed this behavior but find at the unitarity limit 
a condensate fraction slightly smaller ($n_0/(n/2)=0.50$) 
than the mean-field expectation ($n_0/(n/2)=0.66$). 

\section{Time-dependent density functional for 
a Fermi superfluid} 

We propose an action functional $A$ which  
depends on the superfluid order parameter 
$\psi({\bf r},t)$ as follows 
\beq 
A = \int dt \; d^3{\bf r} \; \left\{ 
i\hbar \; \psi^* \partial_t \psi + {c\, \hbar^2 \over 2m} 
\psi^* \nabla^2 \psi - U |\psi|^2 
- {\cal E}(|\psi|^2)|\psi|^2 \right\}
. 
\eeq
The term ${\cal E}$ is the bulk energy per particle 
of the system, which is a function of the number density 
$n({\bf r},t)=|\psi({\bf r},t)|^2$. The Laplacian term 
${c\, \hbar^2 \over 2m} \psi^* \nabla^2 \psi$ 
accounts for corrections to the kinetic energy 
due to spatial variations. In the BCS regime, where
the Fermi gas is weakly interacting, the Laplacian term is 
phenomenological and it is called 
{\it von Weizs\"acker correction} \cite{von}. 
In the BEC regime, where the gas of molecules is Bose condensed, 
the Laplacian term is due to the symmetry-breaking of the 
bosonic field operator and it is 
referred to as {\it quantum pressure}. 
Note that in the deep BEC regime our action functional 
reduces to the Gross-Pitaevskii action functional 
\cite{sala1}. 
In our calculations we set the numerical coefficient $c$ 
of the gradient term equal to unity ($c=1$), to obtain
the correct quantum-pressure term in the BEC regime. 
In the BCS regime, a better phenomenological 
choice for the parameter $c$ could be $c=1/3$ 
as suggested by Tosi {\it et al.} \cite{tosi1,tosi2}, or 
$c=1/36$ as suggested by Zaremba and Tso \cite{zaremba}.
\par
For the initial confining trap,
we consider an axially symmmetric harmonic potential  
\beq
U({\bf r},t) = {m \over 2} 
\left[ {\bar \omega}_{\rho}(t)^2 (x^2 + y^2) + 
{\bar \omega}_z(t)^2 z^2 \right]
,  
\eeq
where ${\bar \omega}_j(t)=\omega_j\Theta(-t)$, with 
$j=1,2,3={\rho},{\rho},z$ and $\Theta(t)$ the step function, 
so that, after the external trap is switched off at $t>0$, 
the Fermi cloud performs a free expansion. 
The Euler-Lagrange equation for the superfluid 
order parameter $\psi({\bf r},t)$ 
is obtained by minimizing the action functional 
$A$. This leads to a time-dependent 
nonlinear Schr\"odinger equation (TDNLSE):  
\beq \label{TDNLSE}
i\hbar \; \partial_t \psi = \left( -{\hbar^2 \over 2m} \nabla^2 
+ U + \mu(|\psi|^2) \right) \psi  \;. 
\eeq
The nonlinear term $\mu$ is the bulk chemical potential 
of the system given by the MC EOS or the EBCS EOS. 
As noted previously, in the deep BEC regime this TDNLSE reduces 
to the familiar Gross-Pitaevskii equation. 
From the TDNLSE one deduces the Landau's hydrodynamics 
equations of superfluids at zero temperature by setting 
$\psi({\bf r},t)= \sqrt{n({\bf r},t)} 
e^{i S({\bf r},t)}$, 
${\bf v}({\bf r},t)= {\hbar \over m} \nabla S({\bf r},t)$, 
and neglecting the
term $(-\hbar^2 \nabla^2 \sqrt{n})/(2m\sqrt{n})$,
which would vanish in the uniform regime.
These hydrodynamics equations are 
\beqa
\partial_t n + {\bf \nabla} \cdot (n {\bf v}) &=& 0 \; , 
\\
m \; \partial_t {\bf v}  + 
\nabla \left( \mu(n) + U({\bf r},t) 
+ {1\over 2} m v^2 \right) &=& 0 \; .  
\label{landau-2}
\eeqa 
These superfluid 
equations differ from the hydrodynamic equations of 
a normal fluid in the superfluid 
velocity field being irrotational, i.e. 
${\bf \nabla} \wedge {\bf v}=0$, so that the vorticity term 
${\bf v}\wedge ({\bf \nabla} \wedge {\bf v})$ does not 
appear in Eq. (\ref{landau-2}). 
\par
By using the superfluid hydrodynamics equations,  
the stationary state in the trap 
is given by the Thomas-Fermi profile 
$n_0({\bf r}) = \mu^{-1}\left( \bar{\mu} 
- U({\bf r},0) \right)$. 
Here $\bar{\mu}$, the chemical potential of the 
inhomogeneous system, is fixed by the normalization condition 
$N = \int d^3{\bf r} \; n_0({\bf r})$, 
where $N$ is the number of Fermi atoms. 
By imposing that the hydrodynamics equations 
satisfy the scaling solutions for the density 
\beq 
n({\bf r},t)= n_0\left( {x\over b_1(t)} ,{y\over b_2(t)} , 
{z\over b_3(t)} \right)/\displaystyle\prod_{k=1}^3 b_k(t) \; , 
\eeq
and the velocity 
\beq 
{\bf v}({\bf r},t) = \left( x {{\dot b}_1(t)\over b_1(t) } , 
y {{\dot b}_2(t)\over b_2(t) }, 
z { {\dot b}_3(t)\over b_3(t) } \right)
, 
\eeq
we obtain three differential equations for the 
scaling variables $b_j(t)$, with $j=1,2,3={\rho},{\rho},z$.  
The dynamics is well approximated by 
evaluating the scaling differential equations at the center 
(${\bf r}={\bf 0}$) of the cloud. 
In this case the variables $b_j(t)$ 
satisfy the local scaling equations (LSE) 
\beq 
\ddot b_j(t) + {\bar \omega}_j(t)^2 \; b_j(t) = 
{\omega_j^2 \over  \displaystyle\prod_{k=1}^3 b_k(t) }\; 
{
{\partial \mu \over \partial n}\left(\bar{n}(t)\right)   
\over 
{\partial \mu \over \partial n}\left(n_0({\bf 0})\right)
} 
\; , 
\eeq
where $\bar{n}(t)={n_0({\bf 0})/\displaystyle
\prod_{k=1}^3 b_k(t) }$. 
Clearly, the LSE depend critically on the EOS $\mu=\mu(n,a_F)$. 
The TDNLSE is solved by using a finite-difference 
Crank-Nicolson predictor-corrector 
method, that we developed to solve the 
time-dependent Gross-Pitaevskii equation 
\cite{sala2}. 
Observe that imaginary-time integration of Eq.~(\ref{TDNLSE}) by
the Crank-Nicolson method generates
the ground-state of Bose condensates in a ring and 
in a double-well \cite{sala3} 
much more accurately than the 
steepest descent method used in the past. 
The simple LSE are instead solved by using 
a standardleap-frog symplectic algorithm, 
succesfully applied to investigate 
the order-to-chaos transition in spatially homogeneous 
field theories \cite{sala4}. 
\par
In Ref.~\cite{nick3} we have compared our time-dependent 
theory with the available experimental data. Moreover,  
we have compared the full TDNLSE with the LSE by using 
both MC EOS and EBCS EOS. We have found that, 
using the same EOS, the TDNLSE gives results always 
very close to the LSE ones. Instead, we have found some 
differences between MC EOS and EBCS EOS.
Figure 1 reports
the aspect ratio and the released energy of a $^6$Li cloud after 
$1.4$ ms expansion from the trap realized at ENS-Paris 
\cite{ens}. In the experiment of Ref. \cite{ens} 
the free expansion 
of $7\cdot 10^4$ cold $^6$Li atoms has been studied 
for different values of $y=(k_Fa_F)^{-1}$ 
around the Feshbach resonance ($y=0$). Unfortunately, 
in this experiment the thermal component is not negligible 
and thus the comparison with the zero-temperature 
theory is not fully 
satisfactory. Figure 1 compares the experimental data of 
Ref. \cite{ens} with the LSE based on both MC and EBCS 
equation of state. 
This figure shows that the aspect ratio predicted by the 
two zero-temperature theories exceeds the finite-temperature 
experimental results. This is not surprising because 
the thermal component tends to hide the hydrodynamic 
expansion of the superflud. 
On the other hand, the released energy 
of the atomic gas is well described by 
the two zero-temperature theories, and the mean-field theory 
seems more accurate, also probably due to the thermal component. 
In Fig. 1 the released energy is defined 
as in Ref. \cite{ens}, i.e.  
on the basis of the rms widths of the cloud. 
By energy conservation, the actual released 
energy is instead given by 
\beq 
E_{\rm rel} = \int d^3{\bf r} \, {\cal E}[n_0({\bf r})] 
\, n_0({\bf r}) \; . 
\label{bibo}
\eeq
It is straightforward to obtain an analytical expression for the 
released energy assuming 
a power-law dependence $\mu = C \; n^{\gamma}$ 
for the chemical potential (polytropic equation of state) 
and writing 
$
{\cal E}[n_0({\bf r})] \simeq {3\over 5} \mu[n_0({\bf r})] = 
{3\over 5} 
{\bar{\mu}\over n_0(0)^{\gamma} } n_0({\bf r})^{\gamma}  
$
where $\gamma$ is the effective polytropic index,  
obtained as the logarithmic derivative of the chemical 
potential $\mu$ \cite{nick1}, namely
\beq 
\gamma (y) = {n \over \mu} {\partial \mu \over \partial n} = 
{  {2 \over 3} f(y) - 
{2y \over 5} f'(y) + {y^2 \over 15}f''(y) 
\over f(y) - {y \over 5} f'(y) } \; . 
\eeq
In this way one finds the simple approximate formula 
\beq \label{approxrelen}
E_{\rm rel}= {3\over 5} N \epsilon_F \, 
{2(1+\gamma(y)) \over 2 + 5 \gamma(y)} \, f(y) \; ,  
\eeq 
where $\epsilon_F=\hbar^2k_F({\bf 0})^2/(2m)$ 
is the Fermi chemical potential at the center of the trap, 
with $k_F({\bf 0})=\left(3\pi^2 n_0({\bf 0})\right)^{1/3}$.  
Fig. 2 shows that this simple approximate
formula which neglects all
details of the initial aspect ratio produces fair
semi-quantitative agreement, in particular 
in the BEC regime, with the actual released energy 
obtained by solving numerically Eq.~(\ref{bibo}).
\par
During the free expansion of the cloud 
the aspect ratio in the BCS regime ($y\ll -1$) 
is measurably different from the one of the BEC 
regime ($y\gg 1$). 
In Ref.~\cite{nick3} we have predicted an interesting effect:  
starting with the same aspect ratio of the cloud,  
at small times ($t\omega_H \lesssim 3$) 
the aspect ratio is larger 
in the BCS region; at intermediate times 
($t\omega_H \simeq 4$) the aspect ratio is enhanced 
close to the unitarity limit ($y=0$); 
eventually at large times ($t\omega_H \gtrsim 5$) 
the aspect ratio becomes larger in the BEC region. 
Here $\omega_H=(\omega_{\bot}^2\omega_z)^{1/3}$ is the geometric 
average of the trapping frequencies. 
This prediction is based on the numerical simulation 
of the LSE shown in Fig. 2, where we plot the aspect ratio of 
the expanding cloud as a function of the inverse interaction 
parameter $y=1/(k_Fa_F)$ 
at successive time intervals. At $t=0$ the aspect ratio 
equals the trap anisotropy $\lambda=0.34$. 
Of course the detailed sequence of deformations depends on 
the experimental conditions and in particular on 
the initial anisotropy, but the qualitative trend of an 
initially faster reversal on the BCS side, later suppressed 
by the BEC gas, is predicted for the expansion of any 
initially cigar-shaped interacting fermionic cloud. 

\section{Conclusions} 

We have shown that the free expansion of a 
Fermi superfluid in the BCS-BEC crossover, that we simulate in 
a hydrodynamic scheme at zero temperature, 
reveals interesting features. 
We have found that the 
Monte-Carlo equation of state and the 
mean-field equation of state give similar 
results for the free expansion 
of a two-spin Fermi gas. The two theories are in reasonable 
agreement with the experimental data, which, 
however, are affected by 
the presence of a thermal component. 
Our Monte-Carlo equation of state and time-dependent 
density functonal can be used to study many other 
interesting properties; for instance, the 
collective oscillations of the Fermi cloud 
\cite{stringari,minguzzi,nick1}, 
the Fermi-Bose mixtures across a Feshbach resonance of 
the Fermi-Fermi scattering length, 
and nonlinear effects like Bose-Fermi solitons 
and shock waves. Finally, we observe that new experimental 
data on collective oscillations 
\cite{grimm} suggest that the Monte-Carlo equation of state 
is more reliable than the mean-field equation of state.

\newpage

\begin{figure}
\centerline{\psfig{file=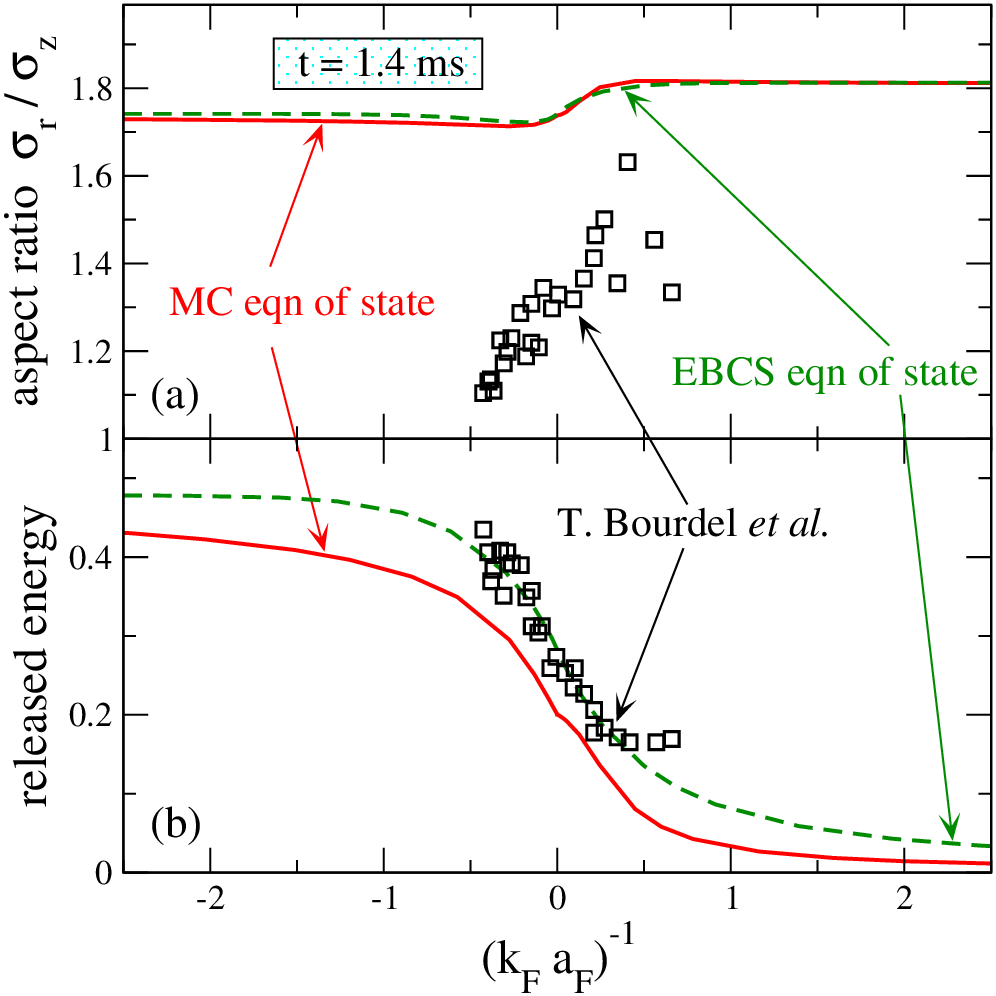,height=5.5in}} 
{\small FIG. 1: Cloud of $N=7\cdot 10^4$ $^6$Li atoms after 
$1.4$ ms expansion from the trap realized 
at ENS-Paris \cite{ens} 
with anisotrpy $\lambda=\omega_z/\omega_{\rho}=0.34$. 
Squares: experimental data. 
Solid lines: numerical simulation with LSE and MC EOS;
dashed lines: numerical simulation with LSE and EBCS EOS.
The released energy is normalized as $E_{\rm rel}/(N\epsilon_F)$.
}
\end{figure}

\newpage 

\begin{figure}
\centerline{\psfig{file=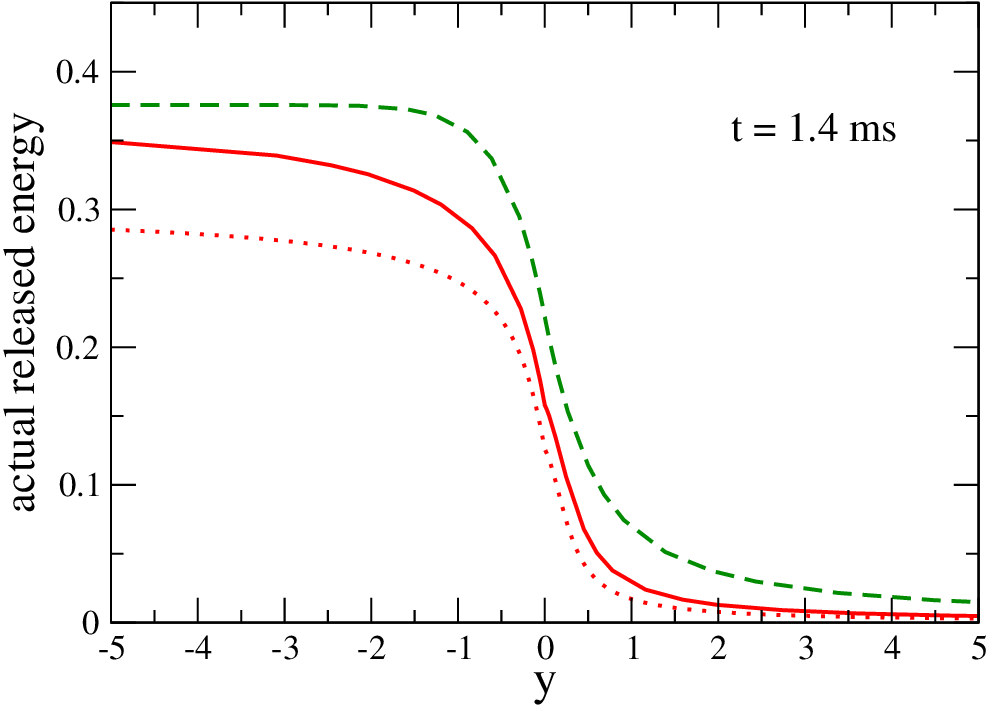,height=4.5in}} 
{\small FIG. 2: Comparison of the approximate
expression of Eq.~(\ref{approxrelen}) (dotted line) to
the actual released energy 
defined in Eq.~(\ref{bibo}) for the conditions
of the ENS-Paris experiment \cite{ens} (initial anisotrpy
$\lambda=\omega_z/\omega_{\rho}=0.34$,
$N=7\cdot 10^4$) (solid line).
Both calculations assume the MC EOS.
The actual released energy based on Eq.~(\ref{bibo})
and the EBCS EOS is also reported (dashed line).}
\end{figure}

\newpage

\begin{figure}
\centerline{\psfig{file=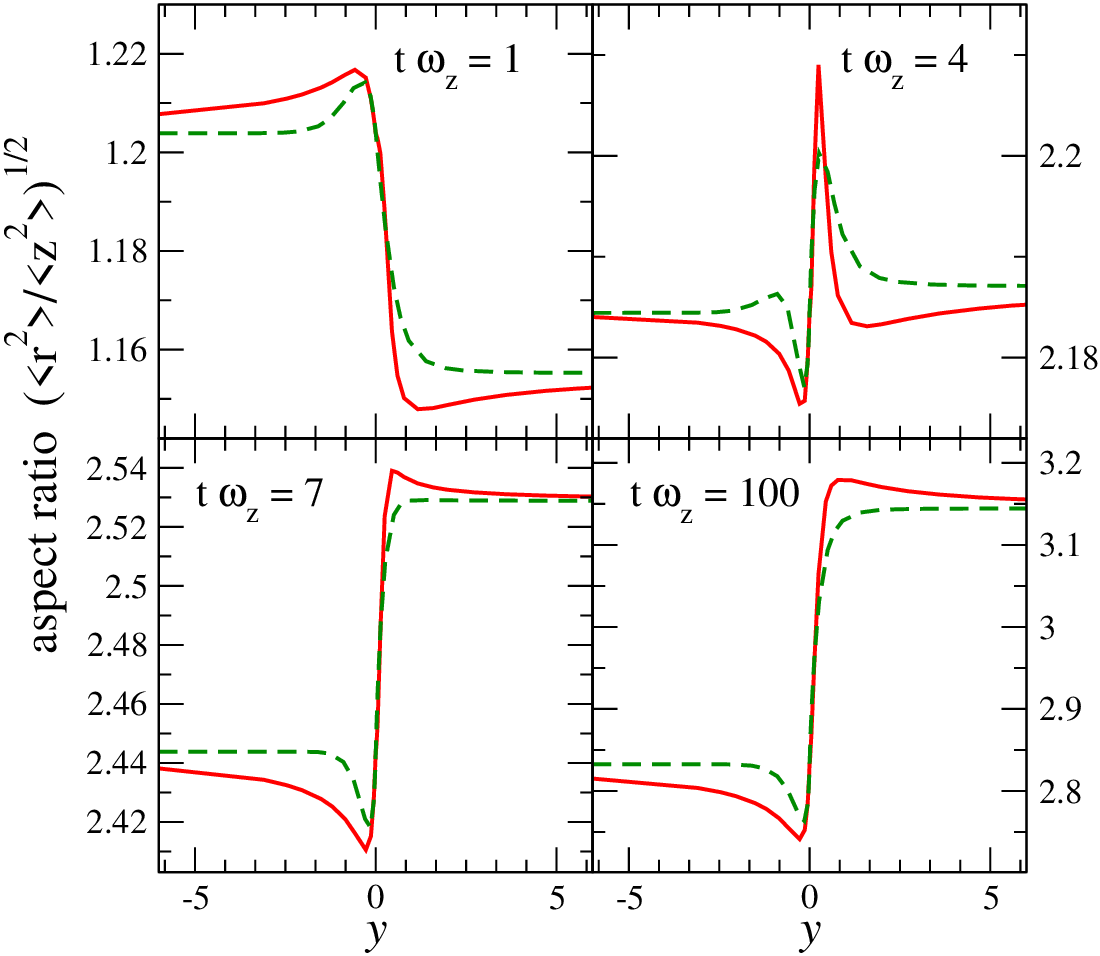,height=5.5in}} 
{\small FIG. 3: Four successive frames of the 
aspect ratio of the $^6$Li Fermi cloud 
as a function of $y=(k_Fa_F)^{-1}$. 
At $t=0$ the Fermi cloud is cigar-shaped with a 
constant aspect ratio equal to the initial 
trap anisotropy $\lambda=\omega_z/\omega_{\rho}=0.34$. 
Solid lines: numerical simulation with LSE and MC EOS; 
dashed lines: numerical simulation with LSE and EBCS EOS.} 
\end{figure}

\end{document}